\documentclass[twoside,11pt]{article}
\usepackage{jmlr2e}
\usepackage{blindtext}
\usepackage{float}
\usepackage{lastpage}
\ShortHeadings{SCORE-SET}{Begari}
\firstpageno{1}

\begin{document}
\title{SCORE-SET: A dataset of GuitarPro files for Music Phrase Generation and Sequence Learning}

\author{\name Vishakh Begari \email vibedjentle@gmail.com \\
       \addr Independent Researcher\\
       Bavaria, Germany\\
       ORCID: \href{https://orcid.org/0009-0008-9399-8581}{0009-0008-9399-8581}}

\editor{Johnny Silverhand}

\maketitle

\begin{abstract}%   <- trailing '%' for backward compatibility of .sty file
A curated dataset of Guitar Pro tablature files (.gp5 format), 
tailored for tasks involving guitar music generation, sequence modeling, and 
performance-aware learning is provided. 
The dataset is derived from MIDI notes in \cite{hawthorne2018enabling} and 
\cite{kong2022giantmidipianolargescalemididataset} which have been 
adapted into rhythm guitar tracks. These tracks are 
further processed to include a variety of expression settings typical 
of guitar performance, such as bends, slides, vibrato, and palm muting, 
to better reflect the nuances of real-world guitar playing. Dataset available at 
\cite{SCORESET}.
\end{abstract}


\section{Introduction}
Advancements in machine learning have led to significant progress in the 
field of automatic music generation, particularly with symbolic representations 
such as MIDI. While datasets like 
\cite{hawthorne2018enabling} 
\cite{7952261} 
\cite{thickstun2017learningfeaturesmusicscratch} 
\cite{bertinmahieux-2011-million} 
\cite{peracha2022jsfakechoralessynthetic}
\cite{bradshaw2025ariamididatasetpianomidi}
\cite{kong2022giantmidipianolargescalemididataset}
have enabled research in mostly piano music generation, there remains a lack of 
large-scale, high-quality resources tailored specifically to the guitar, a highly 
expressive and technically diverse instrument.

Guitar music presents unique challenges for 
modeling due to its polyphonic nature, alternate tunings, and rich 
expressive techniques (e.g., bends, slides, palm muting). Existing symbolic 
music datasets often lack this level of nuance, limiting the development of 
models capable of learning and generating realistic guitar performance.

To address this gap, curated dataset of Guitar Pro tablature files (.gp5 format) is provided 
designed for guitar music generation, sequence modeling, and performance-aware learning. 
The dataset is derived from the MIDI information found in \cite{hawthorne2018enabling} and 
\cite{kong2022giantmidipianolargescalemididataset}, with melodies adapted innto 
rhythm guitar tracks and enriched with expressive elements common in guitar playing.
% Acknowledgements and Disclosure of Funding should go at the end, before appendices and references
\section{SCORE-SET Dataset}
MIDI notes provide information about both pitch and timing, specifying when 
a note is played, its duration, and its musical pitch. 
In the context of guitar tablature, the pitch is mapped to an open string and 
fret position, while the duration is quantized to musical beats. The guitar used is a 6-string instrument tuned to standard E--B--G--D--A--E. 
Both single notes and chords are automatically encoded along with their corresponding beat durations.

To begin with, an overview of articulations to be used in 
the dataset and their tablature is provided. These are deemed essential for capturing the expressive nuances of 
guitar performance.
\subsection{Expressions}
Accentuation in playing refer to emphasising specific notes or rhythms to create dynamics and expression in music. 
\subsubsection{Palm mute}
Palm mute - A technique of lightly resting the edge of palm on the strings near bridge while plucking or strumming.
\subsubsection{Bends}
A bend involves pushing or pulling or pushing a string laterally across the fretboard to raise its pitch.
\begin{figure}[htbp]
  \centering
\includegraphics[trim={40 980 690 100}, clip, width=0.135\linewidth]{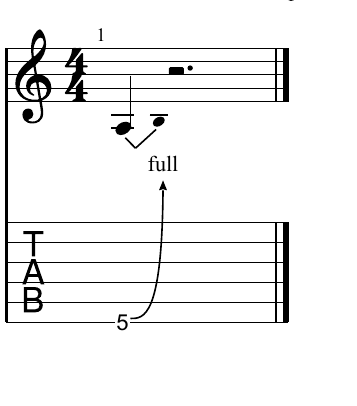}
\includegraphics[trim={40 980 690 100}, clip, width=0.135\linewidth]{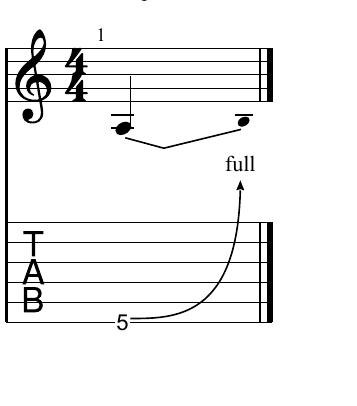}
\includegraphics[trim={40 980 690 100}, clip, width=0.135\linewidth]{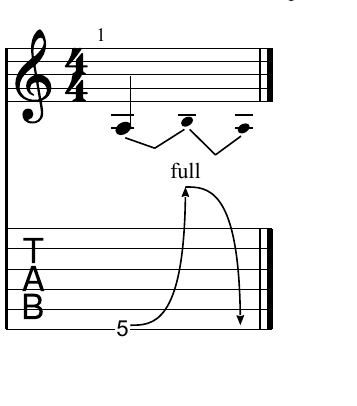}
\includegraphics[trim={40 980 690 100}, clip, width=0.135\linewidth]{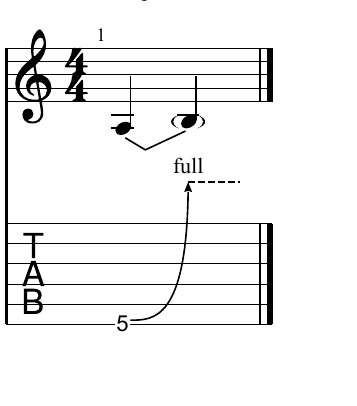}
\includegraphics[trim={40 980 690 100}, clip, width=0.135\linewidth]{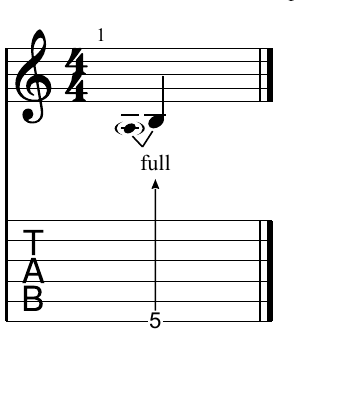}
\includegraphics[trim={40 980 690 100}, clip, width=0.135\linewidth]{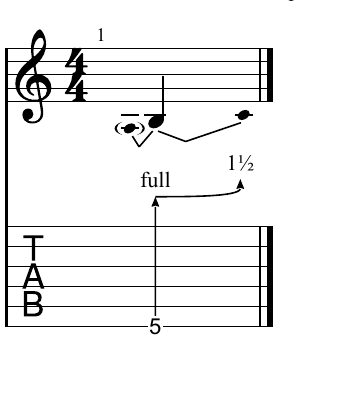}
\includegraphics[trim={40 980 690 100}, clip, width=0.135\linewidth]{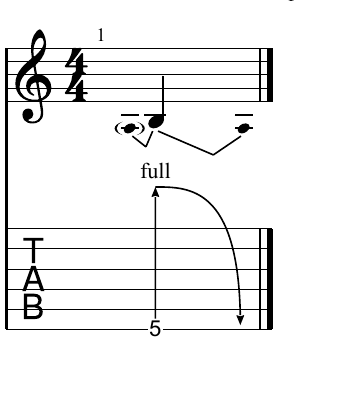}
\includegraphics[clip, width=0.135\linewidth]{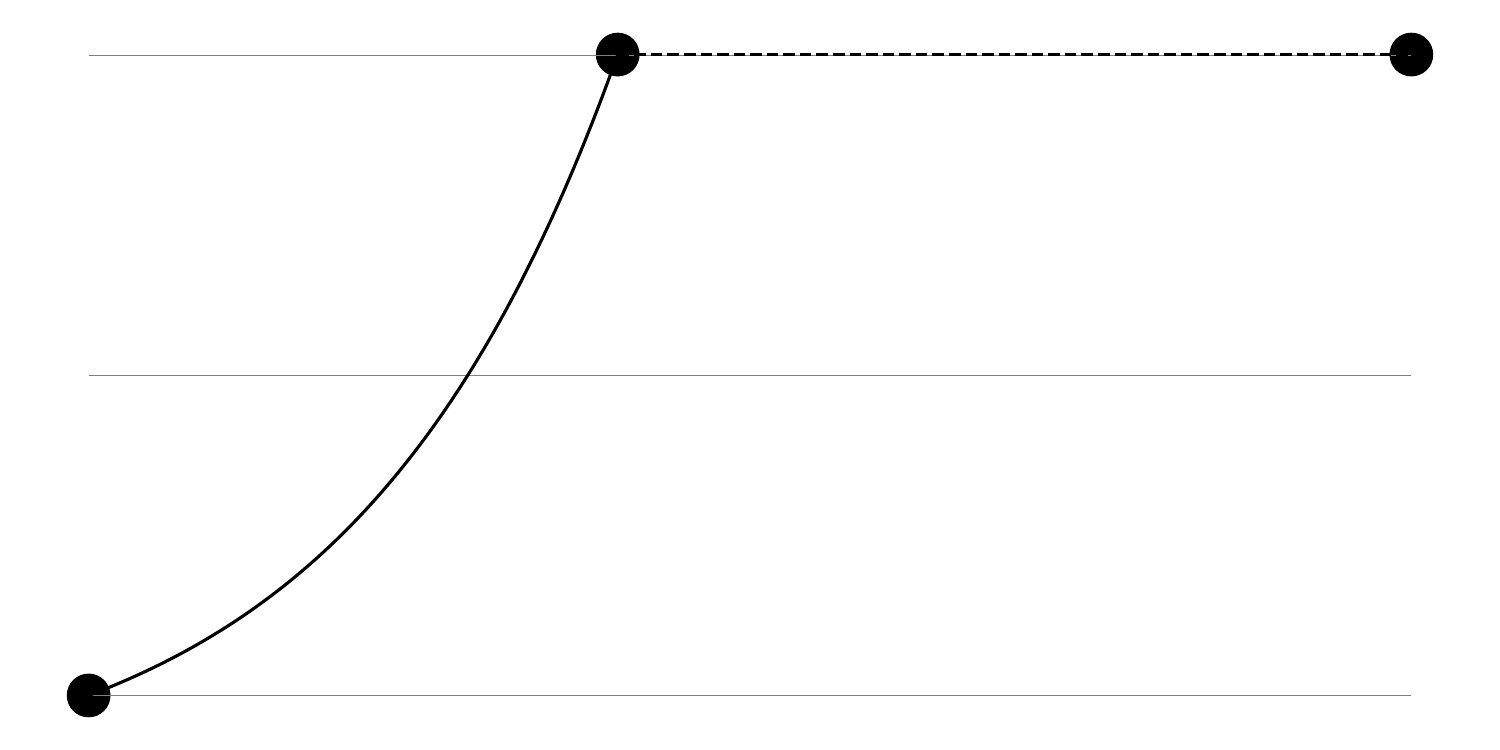}
\includegraphics[clip, width=0.135\linewidth]{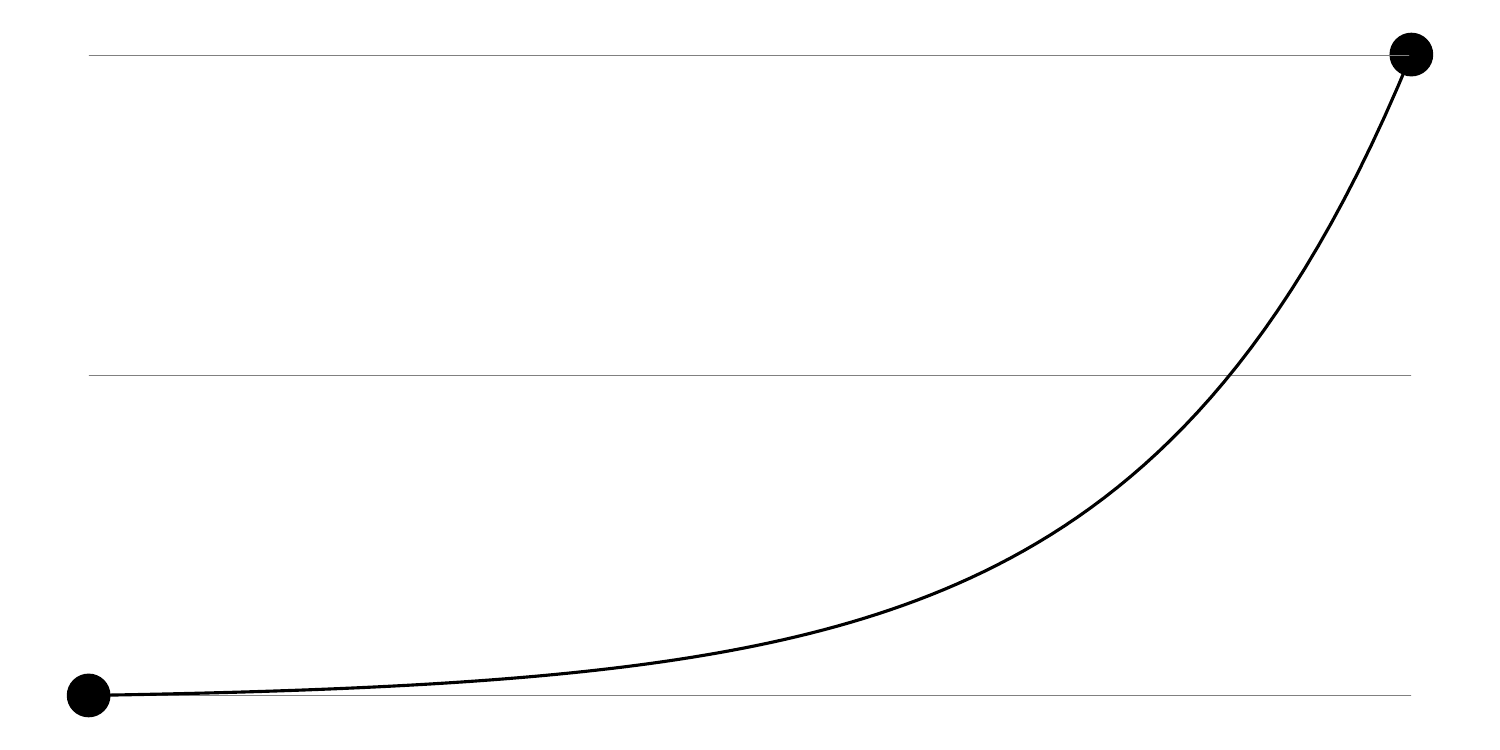}
\includegraphics[clip, width=0.135\linewidth]{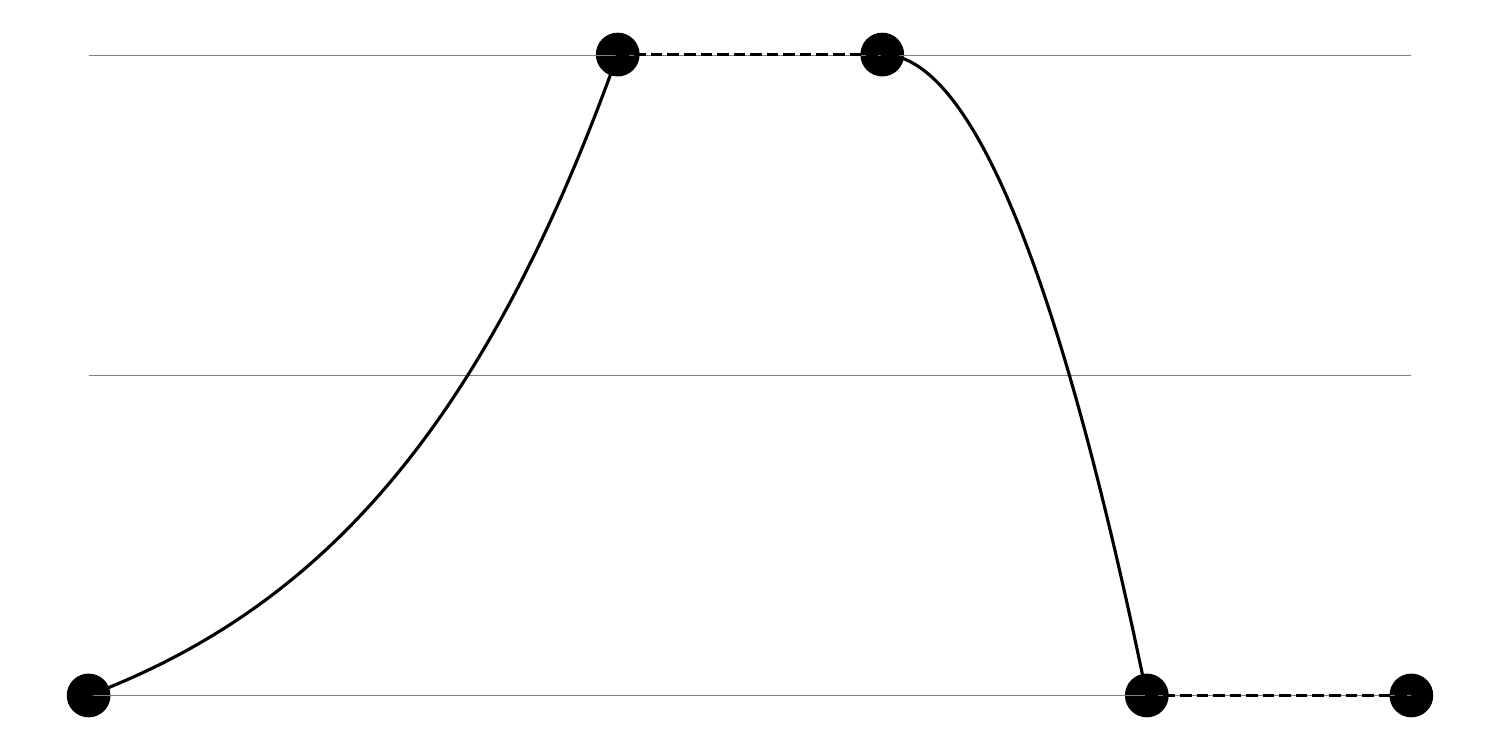}
\includegraphics[clip, width=0.135\linewidth]{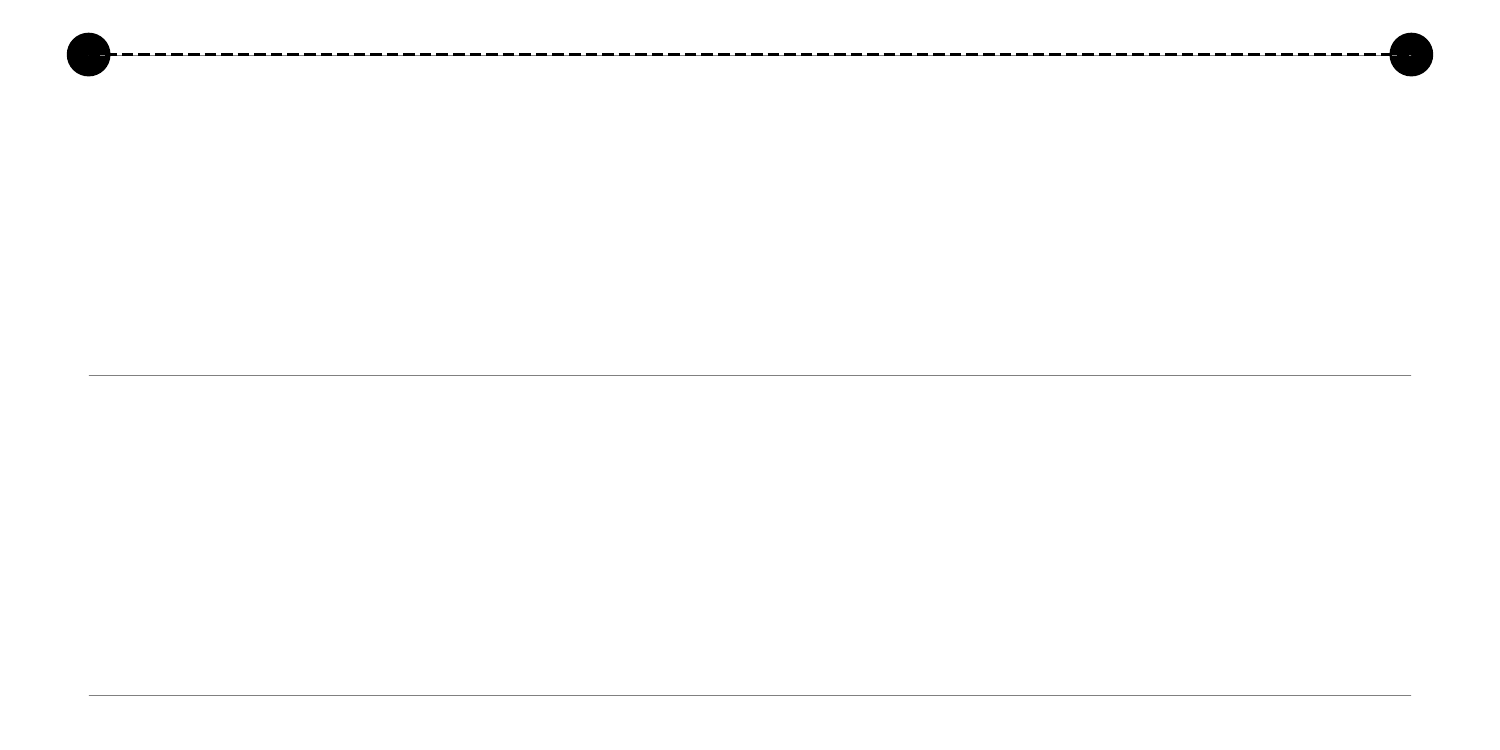}
\includegraphics[clip, width=0.135\linewidth]{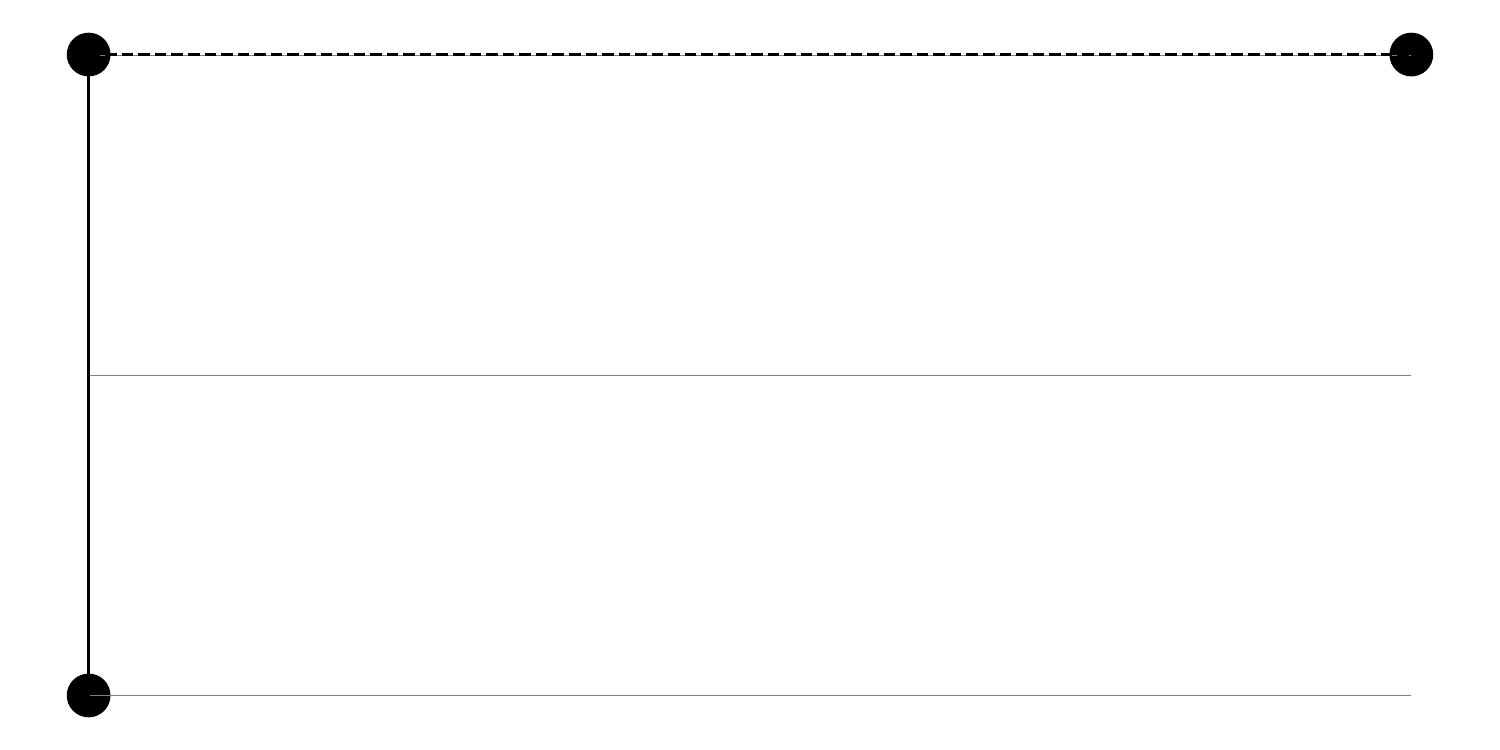}
\includegraphics[clip, width=0.135\linewidth]{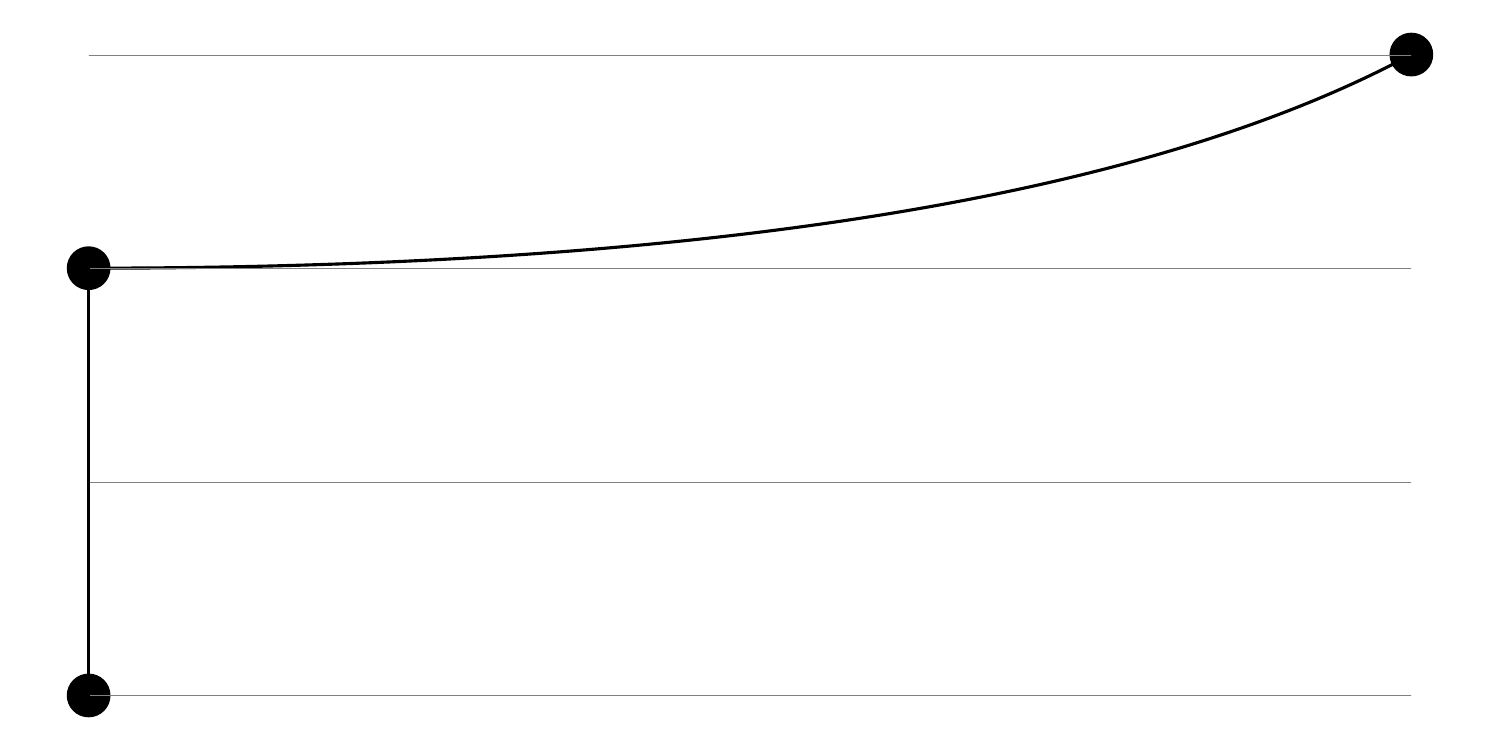}
\includegraphics[clip, width=0.135\linewidth]{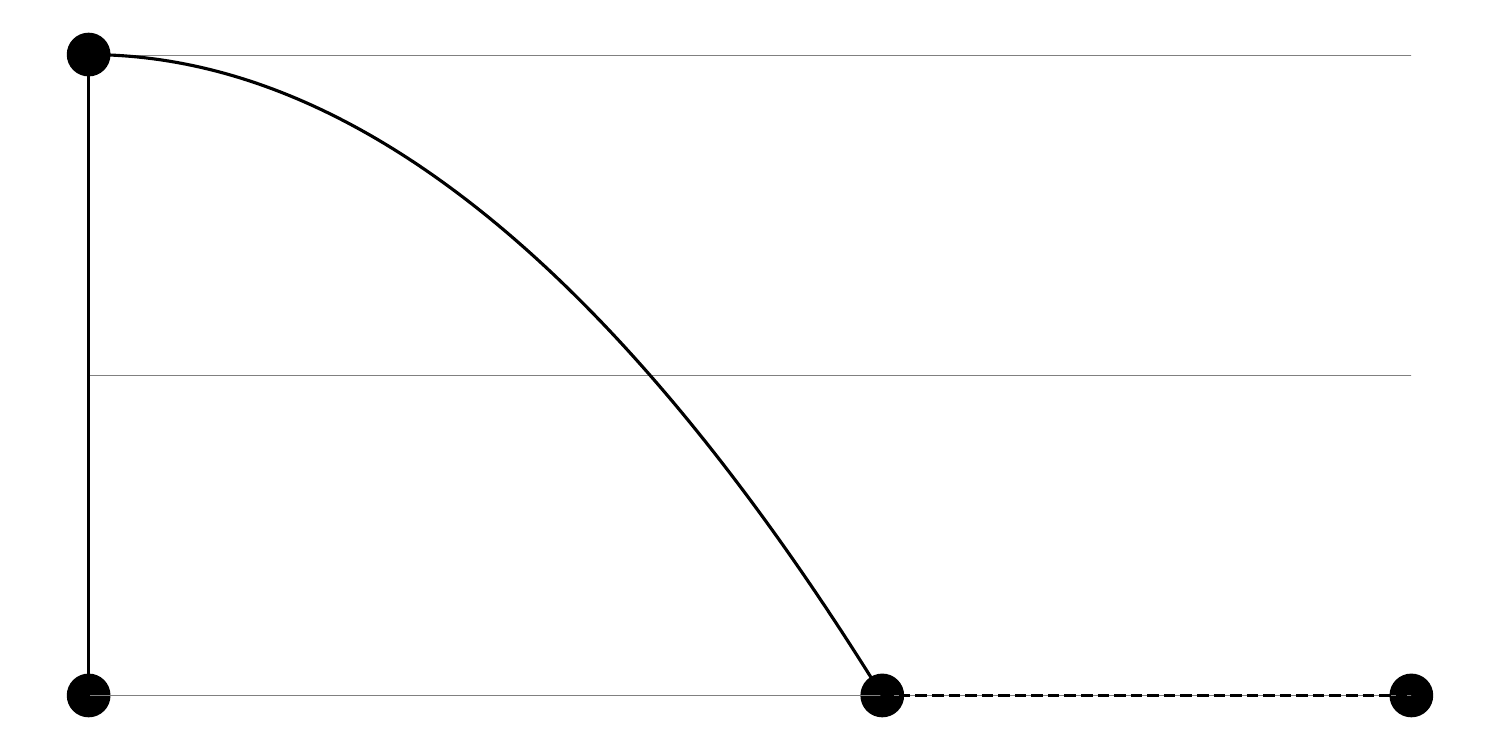}
\caption{Bend types and their pitch variations. Horizontal line represents a semitone.}
  \label{fig:bend}
\end{figure}
\subsubsection{Tremolo bar}
A tremolo bar or whammy bar is a device attached to the bridge of a guitar that 
allows bending the pitch of notes by pushing or pulling on the bar.
\begin{figure}[H]
  \centering
\includegraphics[trim={40 980 690 100}, clip, width=0.195\linewidth]{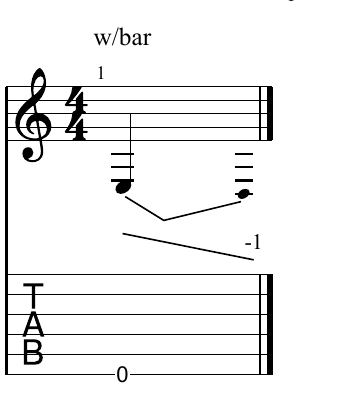}
\includegraphics[trim={40 980 690 100}, clip, width=0.195\linewidth]{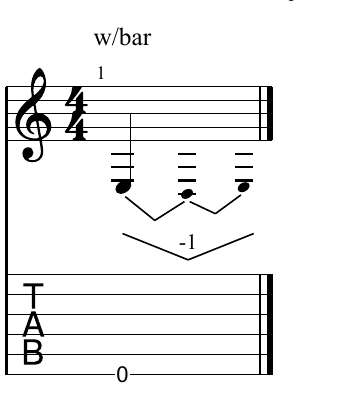}
\includegraphics[trim={40 980 690 100}, clip, width=0.195\linewidth]{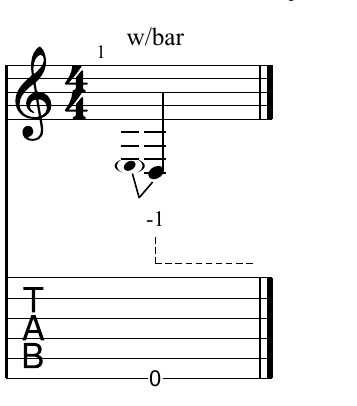}
\includegraphics[trim={40 980 690 100}, clip, width=0.195\linewidth]{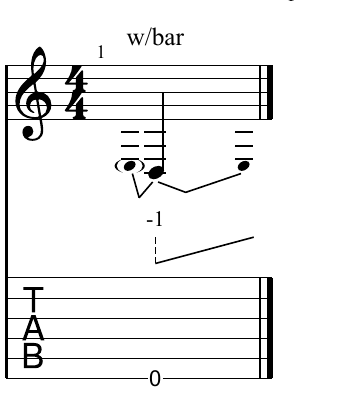}
\caption{Tremlo types and their pitch variations.}
  \label{fig:trem}
\end{figure}
\subsubsection{Slide}
Smoothly moving up or down the fretboard without lifting off on the string, creating a seamless transition between notes.
\begin{figure}[ht]
  \centering
\includegraphics[trim={40 1040 690 100}, clip, width=0.16\linewidth]{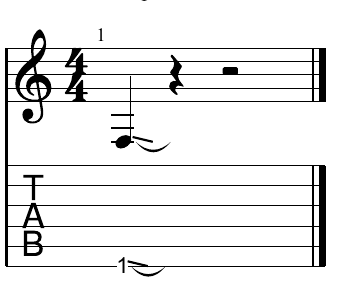}
\includegraphics[trim={40 1040 690 100}, clip, width=0.16\linewidth]{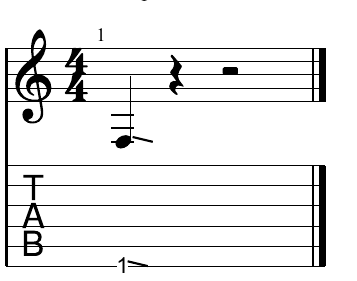}
\includegraphics[trim={40 1040 690 100}, clip, width=0.16\linewidth]{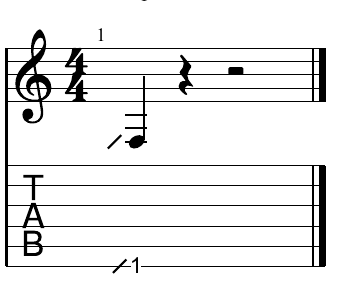}
\includegraphics[trim={40 1040 690 100}, clip, width=0.16\linewidth]{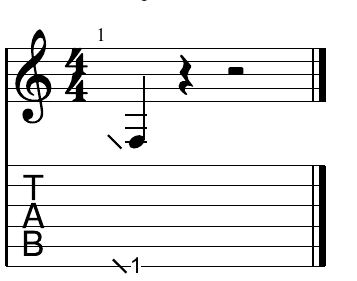}
\includegraphics[trim={40 1040 690 100}, clip, width=0.16\linewidth]{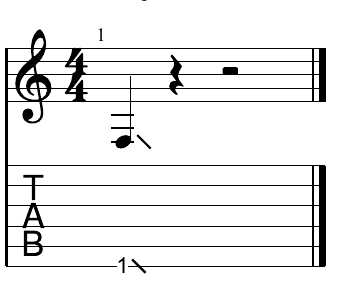}
\includegraphics[trim={40 1040 690 100}, clip, width=0.16\linewidth]{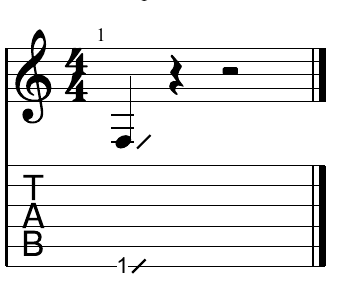}
\caption{Slide types}
\label{fig:slide}
\end{figure}
\subsubsection{Dead note}
Muted note or ghost note - Muting the string to produce a percussive sound.
\subsubsection{Hammer on / pull off}
Allows playing 2 notes in succession without picking the second note. Pressing down onto a higher fret to play a note without picking it, producing smooth legato transition between notes. 
Pulling off a higher fretted note while the lower fret note is still pressed.
\subsubsection{Vibrato}
Involves bending and releasing the string rapidly to oscillate the pitch slightly above and below the target note.
\subsubsection{Harmonic - Natural}
Producing bell-like, chime sound by lightly touching a string at specific points along the fretboard. 
\begin{figure}[H]
  \centering
\includegraphics[trim={40 1040 690 100}, clip, width=0.22\linewidth]{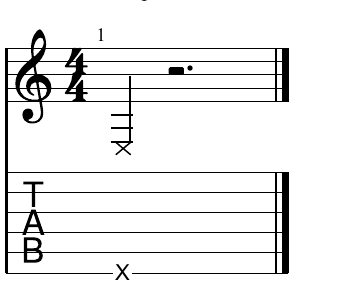}
\includegraphics[trim={25 685 450 70}, clip, width=0.25\linewidth]{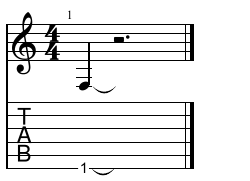}
\includegraphics[trim={40 1020 690 100}, clip, width=0.22\linewidth]{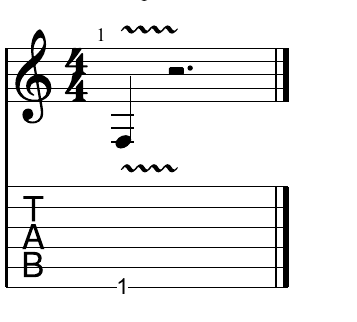}
\includegraphics[trim={40 1040 710 100}, clip, width=0.22\linewidth]{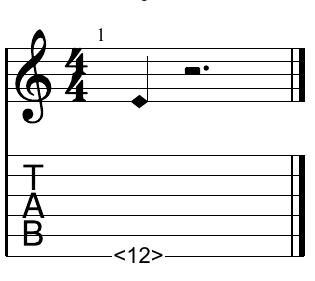}
\caption{From left to right - Dead note, Hammer On, Vibrato, Natural Harmonic}
\label{fig:harmonic}
\end{figure}

\subsection{Quantization}
MIDI formats represent time signatures using ticks, requiring 
quantization techniques to accurately derive rhythmic patterns. For simplicity, 
the minimum note duration was limited to a sixteenth note. Off-beat notes were
quantized to the nearest beat within the bar, and all notes were aligned to 
a 4/4 time signature. In cases where notes extended beyond the standard 
measure length, the measure was dynamically adjusted to accommodate them.

\subsection{String and fret}
Since the original datasets consist of piano-based transcriptions, mapping 
them to guitar fret positions poses a challenge. A custom script was developed to 
translate piano MIDI files into guitar tablature by mapping MIDI 
note pitches to plausible fretboard positions. Each MIDI note was first 
converted to its corresponding pitch class, and then assigned to 
a string-fret combination based on the tuning of the guitar. 
To ensure realistic playability, the algorithm prioritized lower fret 
positions and avoided excessive string skipping. In cases where 
multiple valid positions existed for a single note, a heuristic based on 
finger reach and melodic continuity was used to determine the most natural placement. 
This conversion preserved the timing and duration of the original piano notes while 
rendering them in a format suitable for guitar performance. For chord voicings, the root
note was first fixed, and subsequent notes were assigned to consecutive 
strings. If multiple notes fell on the same string, their pitch was adjusted by 
adding or subtracting octaves until a playable fret position was found on 
different strings.

\section{Statistics}
\begin{figure}[H]
\centering
\includegraphics[clip, width=0.8\linewidth]{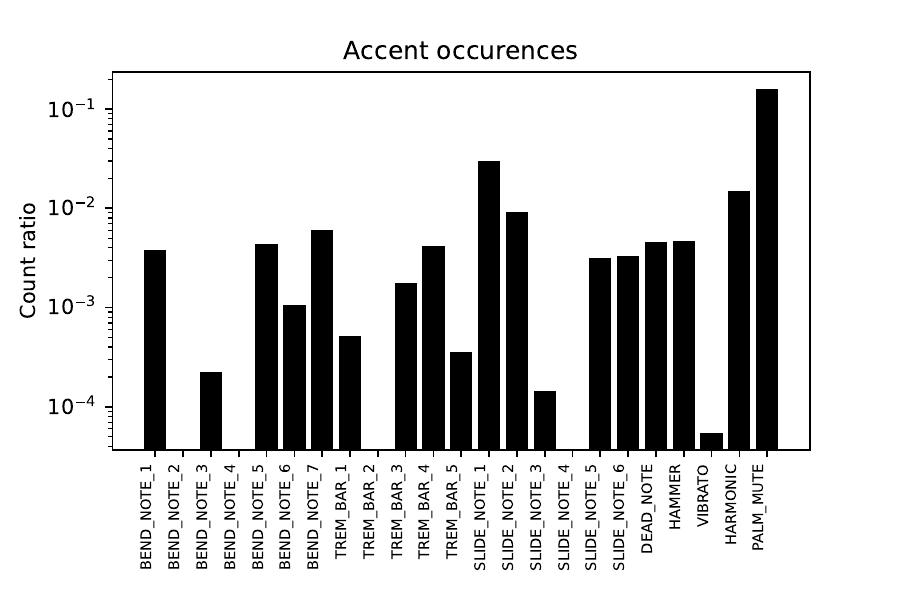}
\caption{Percentages of techniques.}
\label{fig:accent}
\end{figure}
For this study, a dataset of guitar tablatures was manually 
created through transcription of publicly available audio 
recordings of modern metal songs. The transcriptions were 
created by ear and focused specifically on expressive guitar 
techniques. Existing symbolic music datasets often lack 
detailed guitar specific expressions such as palm muting, 
tremolo picking, or slides elements that are crucial to the 
stylistic integrity of modern metal music. Since these 
articulations are rarely annotated in standard MIDI or 
tablature datasets, manual transcription was necessary to 
ensure expressive accuracy and genre fidelity. The resulting 
dataset offers a more nuanced representation of guitar 
performance suited for expression-aware generative modeling.
The expression statistics were calculated over a total of 198656 notes, with approximately 25.39 \% 
of them featuring accents.

\begin{figure}[H]
\centering
\includegraphics[trim={20 620 200 100}, clip, width=0.8\linewidth]{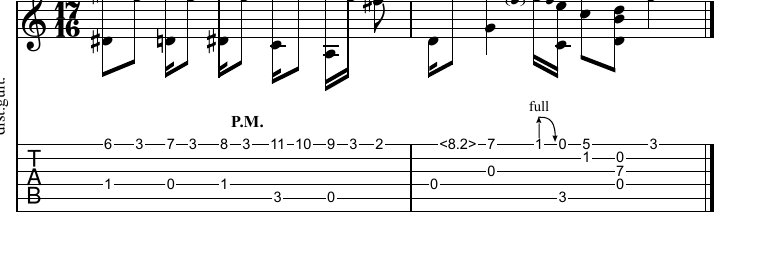}
\caption{Tablature showing palm mute, bend, harmonic and custom measure.}
\label{fig:gpro}
\end{figure}
Accents were added to the original dataset at random, 
following the expression ratios derived from 
Figure~\ref{fig:accent}. Although these percentages were 
based on individual musical notes, tremolo is an exception, 
as it is applied to an entire beat, which may contain multiple 
notes.

After expression augmentation, the sequences were exported 
to .gp5 files using the PyGuitarPro~\cite{PyGuitarPro} library, preserving all 
note level attributes and guitar specific articulations. 
This allowed for compatibility with standard tablature 
software and ensured that the data could be both inspected and 
rendered in a musically meaningful format. The dataset is made available 
at~\cite{SCORESET}.

\acks{This research received no specific grant from any funding agency in the public, commercial, or not-for-profit sectors.}

\vskip 0.2in
\bibliography{SCORESET}

\end{document}